\documentclass[showpacs,preprintnumbers,superscriptaddress,pre]{revtex4}
\usepackage{epsfig}
\usepackage{graphicx}
\usepackage{amsmath,amsfonts,amssymb}
\usepackage{pstricks}
\usepackage{color}
\definecolor{Blue}{rgb}{0.3,0.3,0.9}

\renewcommand{\epsilon}{\varepsilon}
\def\be{\begin{equation}}
\def\ee{\end{equation}}
\newcommand\bea{\begin{eqnarray}}
\newcommand\eea{\end{eqnarray}}

\begin{document}

\title{Exact finite-size corrections for the  spanning-tree model
under different boundary conditions}

\date{\today}

\author{N. Sh. Izmailian}
\email{ab5223@coventry.ac.uk; izmail@yerphi.am}
\affiliation{Applied Mathematics Research Centre, Coventry
University, Coventry CV1 5FB, UK} \affiliation{Yerevan Physics
Institute, Alikhanian Brothers 2, 375036 Yerevan, Armenia}

\author{R. Kenna}
\email{r.kenna@coventry.ac.uk} \affiliation{Applied Mathematics
Research Centre, Coventry University, Coventry CV1 5FB, UK}

\begin{abstract}
We express the partition functions of the spanning tree on finite square lattices under five different sets of boundary conditions in terms of a principal partition function with twisted boundary conditions.
Based on these expressions, we derive the exact asymptotic expansions of the logarithm of the partition function for each case. We have also established several groups of identities relating spanning-tree partition functions for the different boundary conditions. We also explain an apparent discrepancy between logarithmic correction terms in the free energy for a two dimensional spanning tree model with periodic and free boundary conditions and  conformal field theory predictions. We have obtain corner free energy for the spanning tree under free boundary conditions in full agreement with conformal field theory predictions.
\end{abstract}


\maketitle

\section{Introduction}
\label{Introduction}

Systems under various boundary conditions have the same per-site free energy, internal energy, specific heat, etc, in the bulk limit. Finite-size corrections, however, are dependent on the boundaries.
Theories of finite-size effects have been successful in deriving critical and noncritical properties of infinite systems from their finite or partially finite counterparts.
In the quest to improve our understanding of realistic systems of finite extent,
two-dimensional models play crucial roles in statistical
mechanics as they have long served as a testing ground to explore
the general ideas of finite-size scaling under controlled
conditions. Of particular importance in such studies are exact
results where the analysis can be carried out without numerical
errors, the Ising model \cite{ferdinand1969,ising,izm2002}, the
dimer and spanning tree model \cite{dimer,izm2003,tseng2003,izm2014} being the most prominent examples.

In 2002 Ivashkevich, Izmailian, and Hu \cite{izm2002} proposed a
systematic method to compute exact finite-size corrections to the
partition functions and their derivatives of free models on the torus,
including the Ising model, dimer model, and Gaussian model. They
found that the partition functions of all these models can be
written in terms of the partition functions with twisted boundary
conditions $Z_{\alpha,\beta}$ with $(\alpha, \beta) =(1/2,0) ,
(0,1/2)$, and $(1/2,1/2)$. Extending this approach, Izmailian,
Oganesyan, and Hu \cite{izm2003} computed the finite-size
corrections to the free energy for the dimer model on finite
square lattices under five different set of boundary conditions (free,
cylindrical, toroidal, M\"{o}bius strip, and the Klein bottle). They
found that the aspect-ratio dependence of finite-size corrections
is sensitive to boundary conditions and the parity of the number
of lattice sites along the lattice axis. Quite recently, Izmailian
and Kenna \cite{izm2014} have found that the partition functions
of the anisotropic dimer model on the rectangular $(2M - 1) \times
(2N - 1)$ lattice with free and cylindrical boundary conditions
with a single monomer residing on the boundary can be expressed
in terms of a partition function with twisted boundary conditions
$Z_{\alpha,\beta}$ with $(\alpha, \beta) =(0,0)$. Based on these expressions, they derive
the exact asymptotic expansions of the free energy.

In this paper we will consider the spanning-tree model.
Enumeration of spanning trees on a graph is a classical problem of combinatorial graph theory, first considered by Kirchhoff \cite{kirch} in his analysis of electrical networks.
Let $G ={V,E}$ denote a connected graph (without loops) with vertex and
edge sets V and E.
A spanning subgraph of G is a spanning tree $T$
if it has $V-1$ edges with at least one edge incident at each
vertex. The degree of a vertex is the number of edges attached to
it (often denoted coordination number).
According to the Kirchhoff theorem, the number of spanning tree subgraphs on a lattice is
given by the minors of the discrete Laplacian matrix $\Delta$ of this lattice.
The Laplacian matrix $\Delta$ is defined as
\begin{equation}
\Delta=Q - A
\label{delta}
\end{equation}
where $A$ is an ${\cal N} \times {\cal N}$ adjacency matrix, and
${\cal N}$ is the number of  lattice sites. The elements of
matrix $A$ are given by
\begin{eqnarray}
A_{ij} &=& 1, \qquad \mbox{if sites i and j are adjacent}
\nonumber\\
A_{ij}&=& 0, \qquad \mbox{otherwise} \nonumber \end{eqnarray} and
$Q$ is an ${\cal N} \times {\cal N}$ degree matrix of $G$  with
elements
$$Q_{ij} = k_i \delta_{ij},$$
where $k_i$ is the degree of site $i$, and $\delta_{ij}$ is the
Kronecker delta function.
In 2000 Tzeng and Wu \cite{tseng2000}
obtained the closed-form expressions for the spanning tree
generating function for a hypercubic lattice in $d$ dimensions
under free, periodic and a combination of free and periodic
boundary conditions. They also obtained the spanning tree
generating function for a simple quartic net embedded on two
nonorientable surfaces, a M\"{o}bius  strip and the Klein bottle.

In this paper we will express the partition functions of the spanning tree on finite square lattices under five different  sets of boundary conditions (free, cylindrical, toroidal, M\"{o}bius  strip, and Klein bottle) in terms of a principal partition function with twisted boundary conditions. Based on these expressions, we derive
the exact asymptotic expansions of the logarithm of the partition function for all boundary conditions mentioned above. We will show that the exact asymptotic expansion of the free energy for all boundary conditions can be written as
\begin{equation}
f=f_{\rm{bulk}}+\frac{2 f_{1s}}{M}+\frac{2  f_{2s}}{N} + \frac{f_0(z \xi)}{S}+
\sum_{p=1}^\infty \frac{f_{p}(z \xi)}{S^{p+1}}, \label{Fnew}
\end{equation}
where $S=M \times N$ is the area of the lattice, $\xi = N/M$ is
the aspect ratio, $f_{\rm{bulk}}$ is the bulk free energy $f_{1s}$ and
$f_{2s}$ are the surface free energies in the horizontal and vertical directions respectively, along which $x_1$ and $x_2$ are the weights assigned to the edges of the spanning tree with $z=x_1/x_2$, $f_0(z\xi)$ is the leading finite size correction term and $f_{p}(z \xi)$ for $p=1,2,3,...$ are subleading correction terms.

In general, the bulk free energy $f_{\rm{bulk}}$, the surface free energies $f_{1s}$ and $f_{2s}$ and subleading correction terms $f_{p}(z \xi)$ ($p=1,2,3,...$) are non-universal, but the
coefficient $f_0$ is supposed  to be universal \cite{privman1984}.
The value of $f_0$ is known \cite{blote1986} to be
related to the conformal anomaly number $c$ and conformal weights of the underlying
conformal theory. Cardy and Peschel \cite{cardy1988} have shown that corners on the boundary induce a trace anomaly in the stress tensor. They predicted that corner contribution to free energy $f_{\rm{corner}}$ gives rise to a term in $f_0$ equal to
\begin{equation}
f_{\rm{corner}}=-\frac{c}{8}\ln \cal{S}, \label{fcorner}
\end{equation}
where $c$ is the central charge defining the universality class of the system and $\cal{S}$ is the area of the domain.
Later, Kleban and Vassileva \cite{kleban1991} extended the study of the free energy on a rectangle. They further derived a geometry-dependent universal part of the free energy in the rectangular geometry and showed that in addition to corner contribution predicted by Cardy and Peschel \cite{cardy1988}, the term $f_0$ contains also another universal part $f_{u}$ depending on the aspect ratio
\begin{equation}
f_{u}=\frac{c}{4}\ln\left[\eta(q)\eta(q')\right], \label{fu}
\end{equation}
where $\eta(q)=q^{1/24}\prod_{n=1}^{\infty}\left(1-q^n\right)$ is the Dedekind eta function and
$q=\exp{(-2\pi\xi)}$, $q'=\exp{(-2\pi/\xi)}$. Moreover, the term $f_0$ contains also non-universal, geometry-independent constant $f_{\rm{nonuniv}}$. Thus the term $f_0$ can be written as
\begin{equation}
f_0=f_{\rm{univ}}+f_{\rm{nonuniv}}, \label{f0tot}
\end{equation}
where the universal part $f_{\rm{univ}}$ of the free energy in the rectangular geometry can be calculated by conformal field theory methods \cite{kleban1991} and given by
\begin{equation}
f_{\rm{univ}}=f_{\rm{corner}}+f_{u}. \label{funiv}
\end{equation}
while  non-universal part $f_{\rm{nonuniv}}$ of the free energy is not calculable via the conformal field theory methods.

Until now there is little evidence for these predictions from exact solutions or
numerical calculations. Quite recently, an efficient bond propagation algorithm was applied for computing the partition function of the Ising model with free edges and corners in two dimensions on square and triangular lattices \cite{IWX1,IWX2,IWXB}.
They  verify the conformal field theory prediction given by Eq. (\ref{funiv}) with central charge $c = 1/2$. Later  the conformal field theory prediction (\ref{funiv}) was  confirmed \cite{izm2014} for the dimer  model on odd-odd square lattices with one monomer on the boundary, for which the central charge is $c = - 2$. However for another
model in the $c=-2$ universality class, namely the spanning tree model, Duplantier and David \cite{duplantier} found logarithmic correction terms in the free energy for periodic and free boundary conditions, which appeared to contradict the conformal field theory prediction. In what follow we will explain such discrepancy.

In this paper we will show that for the spanning tree on finite square lattices under free boundary conditions $f_0$ contains the universal part $f_{\rm{univ}}$, given by Eq. (\ref{funiv}) and does not contains the term coming from corners of the lattice for periodic, cylindrical, Mobius and Klein bottle boundary conditions. This also confirms the conformal field theory prediction for the corner free energy in models for which the central charge is $c = - 2$. Moreover, the non-universal, geometry-independent constant $f_{\rm{nonuniv}}$, which is not calculable via the method of Kleban and Vassileva
\cite{kleban1991}, is determined
\begin{equation}
f_{\rm{nonuniv}}=-\ln{2}-\frac{1}{4}\ln{(1+z^2)}
+\frac{5}{4}\ln{z}. \label{fnu}
\end{equation}
We have also established several groups of identities relating spanning tree partition functions for the different boundary conditions.

Our objective in this paper is to study the finite-size
properties of a spanning tree on the plane square lattice under
five different set of boundary conditions using the same techniques developed
in Refs. \cite{izm2002}, \cite{izm2003} and \cite{izm2014}. The paper is organized as follows. In Sec. II we
introduce the principal partition functions with twisted boundary conditions
$Z_{\frac{1}{2},\frac{1}{2}}(z,M,N)$ and $Z_{0,0}(z,M,N)$.
In Sec. III we  show how the partition functions of the spanning tree under
different boundary conditions can be expressed in terms of the principal partition
functions with twisted boundary conditions. In Sec. IV we derive
several groups of identities relating spanning-tree partition functions
for the different boundary conditions. In Sec. V we discuss the finite-size corrections of
the spanning-tree model and derive the exact asymptotic expansions
of the logarithm of the partition functions for all five sets of boundary conditions and write down the expansion coefficients up to arbitrary order. Our results are summarized and discussed  in Sec. VI.

\section{Partition function with twisted boundary condition} 
\label{PartitionTwisted}

We will show that the exact partition functions of the anisotropic spanning-tree model on
finite rectangular lattices with free, cylindrical, toroidal, M\"{o}bius-strip
and Klein-bottle boundary conditions can be expressed in terms of
the principal partition functions with twisted boundary conditions
$Z_{\frac{1}{2},\frac{1}{2}}(z,M,N)$
and $Z_{0,0}(z,M,N)$, where
\begin{equation}
Z^2_{\alpha,\beta}(z,M,N)=
\prod_{m=0}^{{M}-1}\prod_{n=0}^{N-1}f\left(\frac{(m+\alpha) \pi}{M},\frac{(n+\beta)\pi}{N}\right), \label{Zab}
\end{equation}
with $(\alpha, \beta) \ne (0, 0)$. Here $f(x,y)$ is given by
\begin{equation}
f(x,y)=4\left(z^2 \sin^2{x}+\sin^2{y}\right).
\label{fxy}
\end{equation}
Since $Z_{0,0}(z,M,N)$ vanishes due to the zero mode at $(m,n)=(0,0)$, therefore, when  $\alpha = \beta=0$ we remove the zero mode and remaining product in Eq. (\ref{Zab}) denote by $Z^2_{0,0}(z,M,N)$.

The general theory for the asymptotic expansion of $Z_{0,0}(z,M,N)$ has been given in \cite{izm2014} and the asymptotic expansion of $Z_{\alpha,\beta}(z,M,N)$ for $(\alpha,\beta) \ne (0,0)$ has been given in \cite{izm2002,izm2003}.

\section{The spanning-tree model}
\label{PartitionDimerSpanning}
The enumeration of spanning trees on a graph or lattice is a problem of long-standing interest
in mathematics and physics and it has been first considered by Kirchhoff in his analysis of electrical networks \cite{kirch}. The enumeration of the weighted spanning trees on the $M \times N$ rectangular lattice concerns with the
evaluation of the tree generating function
\begin{equation}
Z_{M,N}=\sum x_1^{n_h}x_2^{n_v},
\label{span}
\end{equation}
with edge weights $x_1$ and $x_2$ along directions M and N, respectively, and the numbers of edges $n_h$ and $n_v$  in the horizontal and vertical directions and summation runs over all possible spanning tree configurations.

In what follows, we will show that the partition function of the spanning tree on an $M \times N$ lattice
is expressed in terms of the principal partition function with twisted boundary conditions
$Z_{0,0}(z,M,N)$ and $Z_{\frac{1}{2},\frac{1}{2}}(z,M,N)$ only, namely
\begin{eqnarray}
Z_{M,N}^{\rm{Torus}}&=&x_2^{MN-1}\frac{1}{M
N}Z_{0,0}^{2}(z,M,N), \label{ZsptrTorFin}\\
Z_{M,N}^{\rm{Cyl}}&=&x_2^{MN-1}\frac{1}{
2 N z \sinh {\left(M{\rm arcsinh}\,1/z \right)}}Z_{0,0}(z,M,2N), \label{ZsptrCylFin}\\
Z_{M,N}^{\rm{Free}}&=&x_2^{MN-1}\frac{(1+z^2)^{1/4}}{z\sqrt{2M N\sinh {\left(2N{\rm arcsinh}\,z \right)}
\sinh {\left(2M{\rm arcsinh}\,1/z \right)} }}Z_{0,0}^{1/2}(z,2M,2N), \label{ZsptrFreeFin}\\
Z_{M,N}^{\rm{Mob}}&=&x_2^{MN-1} \frac{1}{2 M |\sinh {\left(N{\rm arcsinh}\,z+\frac{i \pi M}{2} \right)}|}
Z_{0,0}(z,M,N)Z_{\frac{1}{2},\frac{1}{2}}(z,M,N), \label{ZsptrMobFin}\\
Z_{M,N}^{\rm{Klein}}&=&x_2^{MN-1}\frac{\coth {\left(N{\rm arcsinh}\,z/a \right)}}{2M}  Z_{0,0}(z,M,2N)
\label{ZsptrKleinFin}
\end{eqnarray}
where $a=1$ for even $M$ and $a=0$ for odd $M$.

\subsection{Partition function of the spanning tree model with toroidal boundary condition}
\label{PartitionSpanning}

For a rectangular $M \times N$ lattice with toroidal boundary conditions, the exact partition function for the weighted spanning trees $Z_{M,N}^{\rm{Torus}}$  is given by \cite{tseng2000}
\begin{eqnarray}
Z_{M,N}^{\rm{Torus}}&=&\frac{1}{M
N}\prod_{m=0}^{M-1}{\prod_{n=0}^{N-1}{\!\!\!\!}^{~^\prime}}
2 \left[\textstyle{ \;x_1 \left(1-\cos\frac{2 m \pi}{M}\right)+ x_2 \left(1- \cos\frac{2 n
\pi }{N}\right) \;}\right], \label{ZsptrTor}
\end{eqnarray}
where the prime on the product denotes the restriction $(m,n) \ne (0,0)$. The partition function can be transformed as
\begin{equation}
Z_{M,N}^{\rm{Torus}}=\frac{x_2^{MN-1}}{M
N}\prod_{m=0}^{M-1}{\prod_{n=0}^{N-1}{\!\!\!\!}^{~^\prime}}
f\left(\frac{m \pi}{M},\frac{n \pi}{N}\right), \label{ZsptrTor1}
\end{equation}
where $z=\sqrt{x_1/x_2}$ and $f(x,y)$ is given by Eq. (\ref{fxy}). Now it is easy to see from Eqs. (\ref{Zab}) and (\ref{ZsptrTor1}) that the partition function of the spanning-tree model on an $M \times N$ rectangular lattice with toroidal boundary conditions can be
expressed in terms of $Z_{0,0}(z,M,N)$ and written in the form given by Eq. (\ref{ZsptrTorFin}).

\subsection{Partition function of the spanning tree model with cylindrical boundary condition}
\label{PartitionSpanningc}

For a rectangular $M \times N$ lattice with cylindrical boundary condition (periodic boundary
conditions in the $M$-direction and free boundary conditions in the
$N$-direction), the exact partition function for the weighted spanning
trees $Z_{M,N}^{\rm{Cyl}}$  is given by \cite{tseng2000}
\begin{eqnarray}
Z_{M,N}^{\rm{Cyl}}&=&\frac{1}{M
N}\prod_{m=0}^{M-1}{\prod_{n=0}^{N-1}{\!\!\!\!}^{~^\prime}}
2\left[\textstyle{ \;x_1 \left(1-\cos\frac{2 m \pi}{M}\right)+ x_2 \left(1- \cos\frac{n
\pi }{N}\right) \;}\right], \label{ZsptrCyl}
\end{eqnarray}
which can be transformed as
\begin{equation}
Z_{M,N}^{\rm{Cyl}}=\frac{x_2^{MN-1}}{M
N}\prod_{m=0}^{M-1}{\prod_{n=0}^{N-1}{\!\!\!\!}^{~^\prime}}
f\left(\frac{m \pi}{M},\frac{n \pi}{2 N}\right), \label{ZsptrCyl1}
\end{equation}
where $f(x,y)$ is given by Eq. (\ref{fxy}). Since
\begin{equation}
f\left(\frac{m \pi}{M},\pi-\frac{n \pi}{2 N}\right)=f\left(\frac{m \pi}{M},\frac{n \pi}{2 N}\right),
\label{ransf1}
\end{equation}
the product over $n$ in Eq. (\ref{ZsptrCyl1}) can be extended up to $2N-1$ and the double product
$\prod_{n=0}^{2N-1}{\prod^{\prime}}_{m=0}^{M-1} f\left(\frac{m \pi}{M},\frac{n \pi}{2 N}\right)$
can be expressed in terms of the $\prod_{n=0}^{N-1}{\prod^{\prime}}_{m=0}^{M-1}f\left(\frac{m \pi}{M},\frac{n \pi}{2 N}\right)$ as
\begin{equation}
\prod_{n=0}^{2N-1}
{\prod_{m=0}^{M-1}{\!\!\!\!}^{{~^\prime}}}
f\left(\frac{m \pi}{M},\frac{n \pi}{2 N}\right)=\frac{
\prod_{m=0}^{M-1}f\left(\frac{m \pi}{M},\frac{\pi}{2}\right)}{\prod_{m=1}^{M-1}f\left(\frac{m \pi}{M},0\right)}
\left[\prod_{n=0}^{N-1}
{\prod_{m=0}^{M-1}{\!\!\!\!}^{~^\prime}}f\left(\frac{m \pi}{M},\frac{n \pi}{2 N}\right)\right]^2. \label{transf2}
\end{equation}
Using the identity \cite{GradshteinRyzhik}
\begin{eqnarray}
 \prod_{m=0}^{M-1}4\textstyle{
\left[~\!\sinh^2\omega + \sin^2\frac{\left(m+\alpha\right) \pi }{M}\right]}&=&4 |\sinh\left(M\; \omega+ i \pi \alpha \right)|^2,
\label{identity1}
\end{eqnarray}
for $\alpha=0$, the products $\prod_{m=0}^{M-1}f\left(\frac{m \pi}{M},\frac{\pi}{2}\right)$ and
$\prod_{m=1}^{M-1}f\left(\frac{m \pi}{M},0\right)$ can be written as
\begin{eqnarray}
\prod_{m=0}^{M-1}f\left(\frac{m \pi}{M},\frac{\pi}{2}\right)&=&\prod_{m=0}^{M-1}4 \left[1+z^2 \sin^2\frac{m \pi }{M}\right]=z^{2M}
\prod_{m=0}^{M-1}4
\left[z^{-2} + \sin^2\frac{m \pi }{M}\right]=4 z^{2 M} \sinh^2{\left(M{\rm arcsinh}\,1/z \right)},
\label{prod1}\\
\prod_{m=1}^{M-1}f\left(\frac{m \pi}{M},0\right) &=& \prod_{m=1}^{M-1}4 z^2 \sin^2\frac{m \pi}{M}= z^{2M  - 2} M^2,
\label{prod2}
\end{eqnarray}
respectively. From Eqs. (\ref{ZsptrCyl1}) - (\ref{transf2}), (\ref{prod1}) and (\ref{prod2}) the
partition function of the spanning tree on cylinder can be expressed as
\begin{equation}
Z_{M,N}^{\rm{Cyl}}=\frac{x_2^{MN-1}}{
2 N z \sinh {\left(M{\rm arcsinh}\,1/z \right)}}\sqrt{\prod_{m=0}^{M-1}
{\prod_{n=0}^{2N-1}{\!\!\!\!}^{~^\prime}}
f\left(\frac{m \pi}{M},\frac{n \pi}{2 N}\right)}. \label{ZsptrCyl2}
\end{equation}
Now it is easy to see from Eqs. (\ref{Zab}) and (\ref{ZsptrCyl2}) that the partition function of the
spanning-tree model on an $M \times N$ rectangular lattice with cylindrical boundary conditions can be
expressed in terms of $Z_{0,0}(z,M,2N)$ and written in the form given by Eq. (\ref{ZsptrCylFin}).

\subsection{Partition function of the spanning tree model with free boundary conditions}
\label{PartitionSpanningf}
Let us now consider a rectangular $M \times N$ lattice with free boundaries. The exact partition function for the
weighted spanning trees $Z_{M,N}^{\rm{Free}}$  is given by \cite{tseng2000}
\begin{equation}
Z_{M,N}^{\rm{Free}}=\frac{1}{M
N}\prod_{m=0}^{M-1}{\prod_{n=0}^{N-1}{\!\!\!\!}^{~^\prime}}
2\left[\textstyle{ \;x_1\left(1-\cos\frac{m \pi}{M}\right)+ x_2\left(1- \cos\frac{n
\pi }{N}\right) \;}\right], \label{ZsptrFree}
\end{equation}
which can be transformed as
\begin{equation}
Z_{M,N}^{\rm{Free}}=\frac{x_2^{MN-1}}{M
N}\prod_{m=0}^{M-1}{\prod_{n=0}^{N-1}{\!\!\!\!}^{~^\prime}}
f\left(\frac{m \pi}{2M},\frac{n \pi}{2 N}\right), \label{ZsptrFree1}
\end{equation}
where $f(x,y)$ is given by Eq. (\ref{fxy}). It is easy to show that $f\left(\pi-\frac{m \pi}{2M},\frac{n \pi}{2 N}\right)=f\left(\frac{m \pi}{2M},\pi-\frac{n \pi}{2 N}\right)=f\left(\frac{m \pi}{2M},\frac{n \pi}{2 N}\right)$.
This allows us to express the double product
$\prod_{n=0}^{2N-1}{\prod^{\prime}}_{m=0}^{2M-1} f\left(\frac{m \pi}{2M},\frac{n \pi}{2 N}\right)$, which is
\begin{equation}
\prod_{n=0}^{2N-1}{\prod_{m=0}^{2M-1}{\!\!\!\!}^{~^\prime}} f\left(\frac{m \pi}{2M},\frac{n \pi}{2 N}\right)=\prod_{n=0}^{2N-1}
{\prod_{m=0}^{2M-1}{\!\!\!\!}^{~^\prime}}4\left[\textstyle{ \;z^2\sin^2\frac{m \pi}{2M}+
\sin^2\frac{n \pi}{2 N} \;}\right]=Z^2_{0,0}(z,2M,2N)\label{phiZ}
\end{equation}
in terms of
$\prod_{n=0}^{N-1}{\prod^{\prime}}_{m=0}^{M-1}f\left(\frac{m \pi}{2M},\frac{n \pi}{2 N}\right)$ through
\begin{equation}
\prod_{n=0}^{2N-1}
{\prod_{m=0}^{2M-1}{\!\!\!\!}^{{~^\prime}}}
f\left(\frac{m \pi}{2M},\frac{n \pi}{2 N}\right)=\frac{f\left(0,\frac{\pi}{2}\right)f\left(\frac{\pi}{2},0\right)}{f\left(\frac{\pi}{2},\frac{\pi}{2}\right)}
\frac{\prod_{m=0}^{2M-1}f\left(\frac{m \pi}{2M},\frac{\pi}{2}\right)\prod_{n=0}^{2N-1}f\left(\frac{\pi}{2},\frac{n \pi}{2 N}\right)}
{\prod_{m=1}^{2M-1}f\left(\frac{m \pi}{2M},0\right)\prod_{n=1}^{2N-1} f\left(0,\frac{n \pi}{2 N}\right)}
\left[\prod_{n=0}^{N-1}
{\prod_{m=0}^{M-1}{\!\!\!\!}^{~^\prime}}f\left(\frac{m \pi}{2M},\frac{n \pi}{2 N}\right)\right]^4. \label{phi1}
\end{equation}
Now with the help of identity given by Eq. (\ref{identity1}) we can find that
\begin{eqnarray}
\prod_{m=0}^{2M-1}f\left(\frac{m \pi}{2M},\frac{\pi}{2}\right)&=&z^{4M}\prod_{m=0}^{2M-1}4
\left[z^{-2} + \sin^2\frac{m \pi }{2M}\right]=4 z^{4 M} \sinh^2{\left(2M{\rm arcsinh}\,1/z \right)},
\label{prod10}\\
\prod_{n=0}^{2N-1}f\left(\frac{\pi}{2},\frac{n \pi}{2 N}\right)&=&\prod_{n=0}^{2N-1}4 \left[z^2+ \sin^2\frac{n \pi }{2N}\right]=
4  \sinh^2{\left(2N{\rm arcsinh}\,z \right)},
\label{prod11}\\
\prod_{m=1}^{2M-1}f\left(\frac{m \pi}{2M},0\right) &=& \prod_{m=1}^{2M-1}4 z^2 \sin^2\frac{m \pi}{2M}=4 z^{4M  - 2} M^2,
\label{prod12}\\
\prod_{n=1}^{2N-1}f\left(0,\frac{n \pi}{2 N}\right) &=& \prod_{n=1}^{2N-1}4\sin^2\frac{n \pi}{2N}= 4 N^2. \label{prod13}
\end{eqnarray}
It is easy to show that
\begin{equation}
\frac{f\left(0,\frac{\pi}{2}\right)f\left(\frac{\pi}{2},0\right)}{f\left(\frac{\pi}{2},\frac{\pi}{2}\right)}=
\frac{4z^2}{1+z^2}.
\label{phi10}
\end{equation}
Now, plugging Eqs. (\ref{prod10}) - (\ref{phi10}) back to Eq. (\ref{phi1}) and using Eq. (\ref{phiZ})
we obtain
\begin{equation}
Z^2_{0,0}(z,2M,2N)=\frac{4 z^4 \sinh^2 {\left(2N{\rm arcsinh}\,z \right)}\sinh^2
{\left(2M{\rm arcsinh}\,1/z \right)}}{(1+z^2)M^2 N^2}
\left[\prod_{n=0}^{N-1}
{\prod_{m=0}^{M-1}{\!\!\!\!}^{~^\prime}}f\left(\frac{m \pi}{2M},\frac{n \pi}{2 N}\right)\right]^4.
\label{phiZ1}
\end{equation}
Finally, from Eqs. (\ref{ZsptrFree1}) and (\ref{phiZ1}), the partition function of
the spanning-tree model on an $M \times N$ rectangular lattice with free boundary conditions can be
written in the form given by Eq. (\ref{ZsptrFreeFin}).

\subsection{Partition function of the spanning tree model with M\"{o}bius  strip boundary condition}
\label{PartitionSpanningm}

For a rectangular $M \times N$ lattice with M\"{o}bius  strip boundary
conditions (with free boundary conditions in the $M$-direction and
twisted boundaries in the $N$-direction), the exact partition function for
the weighted spanning trees $Z_{M,N}^{\rm{Mob}}$  is given by \cite{tseng2000}
\begin{eqnarray}
Z_{M,N}^{\rm{Mob}}&=&\frac{1}{M
N}\prod_{m=0}^{M-1}{\prod_{n=0}^{N-1}{\!\!\!\!}^{~^\prime}}
2\left[\textstyle{ \;x_1 \left(1-\cos\frac{m \pi}{M}\right)+ x_2 \left(1- \cos\frac{4 n+1-(-1)^m
}{2N}\pi\right) \;}\right] \label{ZsptrMob}\\
&=&\frac{x_2^{MN-1}}{M
N}\prod_{m=0}^{M-1}{\prod_{n=0}^{N-1}{\!\!\!\!}^{~^\prime}}
4\left[\textstyle{ \;z^2 \sin^2\frac{m \pi}{2M}+ \sin^2\frac{4n+1-(-1)^m
}{4N}\pi \;}\right]. \label{ZsptrMob1}
\end{eqnarray}
The double product in Eq. (\ref{ZsptrMob1}) can be split into two part by considering even $m$ and odd $m$ separately,
\begin{equation}
Z_{M,N}^{\rm{Mob}}=\frac{x_2^{MN-1}}{M
N}\prod_{m=0}^{\lfloor \frac{M-1}{2} \rfloor}{\prod_{n=0}^{N-1}{\!\!\!\!}^{~^\prime}}
f\left(\frac{m \pi}{M},\frac{n \pi}{N}\right)\prod_{m=0}^{\lfloor \frac{M}{2} \rfloor-1}\prod_{n=0}^{N-1}
f\left(\frac{(m+1/2) \pi}{M},\frac{(n+1/2)\pi}{N}\right), \label{ZsptrMob2}
\end{equation}
where $\lfloor a \rfloor$ is the integer part of $a$ and $f(x,y)$ is given by Eq. (\ref{fxy}).

For even $M$
\begin{eqnarray}
Z_{M,N}^{\rm{Mob}}&=&\frac{x_2^{MN-1}}{M
N}\prod_{m=0}^{\frac{M}{2}-1}{\prod_{n=0}^{N-1}{\!\!\!\!}^{~^\prime}}
f\left(\frac{m \pi}{M},\frac{n \pi}{N}\right)\prod_{m=0}^{\frac{M}{2}-1}\prod_{n=0}^{N-1}
f\left(\frac{(m+1/2) \pi}{M},\frac{(n+1/2)\pi}{N}\right) \label{ZsptrMobeven}\\
&=& \frac{x_2^{MN-1}}{2 M \sinh {\left(N{\rm arcsinh}\,z \right)}}Z_{0,0}(z,M,N)Z_{\frac{1}{2},
\frac{1}{2}}(z,M,N). \label{ZsptrMobeven1}
\end{eqnarray}
Here we extend the product over $m$ in the double products of Eq. (\ref{ZsptrMobeven}) up to $M-1$.  For odd $M$
\begin{eqnarray}
Z_{M,N}^{\rm{Mob}}&=&\frac{x_2^{MN-1}}{M
N}\prod_{m=0}^{\frac{M-1}{2}}{\prod_{n=0}^{N-1}{\!\!\!\!}^{~^\prime}}
f\left(\frac{m \pi}{M},\frac{n \pi}{N}\right)\prod_{m=0}^{\frac{M-1}{2}-1}\prod_{n=0}^{N-1}
f\left(\frac{(m+1/2) \pi}{M},\frac{(n+1/2)\pi}{N}\right)\label{ZsptrMobodd}\\
&=& \frac{x_2^{MN-1}}{2 M  \cosh {\left(N{\rm arcsinh}\,z \right)}}Z_{0,0}(z,M,N)Z_{\frac{1}{2},
\frac{1}{2}}(z,M,N). \label{ZsptrMobodd1}
\end{eqnarray}
Here again we extend the product over $m$ in the double products of Eq. (\ref{ZsptrMobodd}) up to $M-1$.

Thus, we have shown that the partition function of the spanning tree on the $M \times N$ lattice with
M\"{o}bius  boundary condition is expressed in terms of the principal partition function with twisted
boundary conditions $Z_{0,0}(z,M,N)$ and $Z_{\frac{1}{2},\frac{1}{2}}(z,M,N)$ only, and can be written
in the form given by Eq. (\ref{ZsptrMobFin}).

\subsection{Partition function of the spanning tree model with Klein bottle boundary condition}
\label{PartitionSpanningk}

For a rectangular $M \times N$ lattice
with Klein-bottle boundary condition (periodic boundary
conditions in the $M$-direction and twisted boundaries in the $N$-direction),
the exact partition function for the weighted spanning trees
$Z_{M,N}^{\rm{Cyl}}$  depends on the parity of the $M$ and given
by \cite{tseng2000}
\begin{eqnarray}
Z_{M,N}^{\rm{Klein}}&=&\frac{2^{M N -1}}{M
N}\prod_{k=1}^{N-1}x_2\left(1-\cos\frac{2\pi n}{N}\right)\prod_{m=1}^{\frac{M-1}{2}}
\prod_{n=0}^{2N-1}
\left[\textstyle{ \;x_1 \left(1-\cos\frac{2 m \pi}{M}\right)
+ x_2 \left(1- \cos\frac{n \pi
}{N}\right) \;}\right],
\label{ZsptrKleinOdd}
\end{eqnarray}
for odd $M$ and
\begin{eqnarray}
Z_{M,N}^{\rm{Klein}}&=&\frac{2^{M N -1}}{M
N}\prod_{n=1}^{N-1}x_2\left(1-\cos\frac{2\pi n}{N}\right)\prod_{n=0}^{N-1}\left[2x_1+
x_2\left(1-\cos\frac{(2n+1)\pi}{N}\right)\right]\nonumber\\
&\times& \prod_{m=1}^{\frac{M}{2}-1}\prod_{n=0}^{2N-1}
\left[\textstyle{ \;x_1 \left(1-\cos\frac{2 m \pi}{M}\right)+ x_2 \left(1- \cos\frac{n \pi
}{N}\right) \;}\right], \label{ZsptrKleinEven}
\end{eqnarray}
for even $M$. The partition function can be transformed as
\begin{eqnarray}
Z_{M,N}^{\rm{Klein}}&=&\frac{x_2^{MN-1}}{M
N}\prod_{k=1}^{N-1}4\sin^2\frac{\pi n}{N}\prod_{m=1}^{ \frac{M-1}{2}}\prod_{n=0}^{2N-1}
f\left(\frac{m \pi}{M},\frac{n \pi}{2N}\right) \label{ZsptrKleinOdd1}\\
&=&x_2^{MN-1}\frac{N}{M}\prod_{m=1}^{ \frac{M-1}{2}}\prod_{n=0}^{2N-1}
f\left(\frac{m \pi}{M},\frac{n \pi}{2N}\right),\label{ZsptrKleinOdd2}
\end{eqnarray}
for odd $M$ and
\begin{eqnarray}
Z_{M,N}^{\rm{Klein}}&=&\frac{x_2^{MN-1}}{M
N}\prod_{n=1}^{N-1}4\sin^2\frac{\pi n}{N}\prod_{n=0}^{N-1}4\left[z^2+\sin^2\frac{(n+1/2)\pi}{N}\right]
\prod_{m=1}^{\frac{M}{2}-1}\prod_{n=0}^{2N-1}
f\left(\frac{m \pi}{M},\frac{n \pi}{2N}\right) \label{ZsptrKleinEven1}\\
&=&\frac{4 x_2^{MN-1}\cosh^2 {\left(N{\rm arcsinh}\,z \right)}}{M
}\prod_{m=1}^{\frac{M}{2}-1}\prod_{n=0}^{2N-1}
f\left(\frac{m \pi}{M},\frac{n \pi}{2N}\right), \label{ZsptrKleinEven2}
\end{eqnarray}
for even $M$. Here we have used the identity given by Eq. (\ref{identity1}) for $\alpha = 0$ and $1/2$.

Now extending product over $m$ in the double product of Eqs. (\ref{ZsptrKleinOdd2}) and
(\ref{ZsptrKleinEven2}) up to $M-1$ we  obtain
\begin{eqnarray}
Z_{M,N}^{\rm{Klein}} &=&\frac{x_2^{MN-1}}{2M}Z_{0,0}(z,M,2N) \hspace{3cm} \mbox{for odd} \; M,
\label{ZsptrKleinEven3}\\
Z_{M,N}^{\rm{Klein}}&=&x_2^{MN-1}\frac{\coth {\left(N{\rm arcsinh}\,z \right)}}{2M}  Z_{0,0}(z,M,2N)
\hspace{1cm} \mbox{for even} \; M. \label{ZsptrKleinOdd3}
\end{eqnarray}
Thus, we have shown that the partition function of the spanning-tree model on an $M \times N$
rectangular lattice with Klein-bottle boundary conditions is expressed in terms of the principal
partition function with twisted boundary conditions $Z_{0,0}(z,M,2N)$ and can be written in the form
given by Eq. (\ref{ZsptrKleinFin}).

Eqs. (\ref{ZsptrTorFin}) - (\ref{ZsptrKleinFin}) give how the partition functions of  the spanning-tree
model on an $M \times N$ rectangular lattice with different boundary conditions can be expressed in
terms of the principal objects $Z_{0, 0}(z,M,N)$ and $Z_{\frac{1}{2},\frac{1}{2}}(z,M,N)$. In the next
section, based on such results, we will established a group of identities relating spanning-tree
partition functions for the different boundary conditions.

\section{Identities for the Spanning Tree Model}
\label{Symmetry}

From Eqs. (\ref{ZsptrTorFin}) - (\ref{ZsptrFreeFin}) and (\ref{ZsptrKleinFin}) one can see that the partition functions of the spanning tree on $M \times N$ lattices with toroidal, cylindrical, free and Klein bottle boundary conditions are all expressed in terms of the principal objects $Z_{0, 0}(z,{\cal M},{\cal N})$ only. Based on such results, it is easy to establish the following group of identities relating spanning tree partition functions for the different boundary conditions
\begin{eqnarray}
Z_{M,2N}^{Torus}&=&\frac{2 M x_2}{N}\left(Z_{M,N}^{Klein}\right)^2  \hspace{0.9cm} \mbox{for odd} \;
M,
\label{idenKTodd}\\
Z_{M,2N}^{Torus}&=&A_1\left(Z_{M,N}^{Klein}\right)^2  \hspace{1.5cm} \mbox{for even} \; M,
\label{idenKTeven}\\
Z_{M,2N}^{Torus}&=&A_2 \left(Z_{M,N}^{Cyl}\right)^2,
\label{idenTC}\\
Z_{2M,2N}^{Torus}&=&A_3 \left(Z_{M,N}^{Free}\right)^4,
\label{idenTF}
\end{eqnarray}
where the coefficients $A_1$ and $A_2$ are given by
\begin{eqnarray}
A_1&=&\frac{2 M x_2}{N \coth^2 {\left(N{\rm arcsinh}\,z \right)} },\label{A1}\\
A_2&=& \frac{2 N x_1 \sinh^2 {\left(M{\rm arcsinh}\,1/z \right)}}{M},\label{A2}\\
A_3&=& \frac{M N x_2^3 z^4 \sinh^2 {\left(2M{\rm arcsinh}\,1/z \right)}\sinh^2 {\left(2N{\rm arcsinh}\,z
\right)}}{1+z^2}.
\label{A3}
\end{eqnarray}
Thus we have established a group of identities relating spanning-tree partition functions for the
toroidal, cylindrical, free and Klein bottle boundary conditions (see Eqs. (\ref{idenKTodd}) - (\ref{idenTF})).

\section{Asymptotic expansion of free energy}
\label{free energy}

In  section II we have expressed the partition functions of the spanning tree on finite square
lattices under five different boundary conditions (free, cylindrical, toroidal, M\"{o}bius  strip, and
Klein bottle) in terms of a principal partition function with twisted boundary conditions
$Z_{0,0}(z,M,N)$ and $Z_{\frac{1}{2},\frac{1}{2}}(z,M,N)$ only.
Based on such results, one can use the exact asymptotic expansions of $Z_{0,0}(z,M,N)$ and
$Z_{\frac{1}{2},\frac{1}{2}}(z,M,N)$ given in papers \cite{izm2014} and \cite{izm2002} to derive the
exact asymptotic expansions of the free energy of the spanning tree $F = - \ln Z$ for all boundary
conditions mentioned above in terms of the
Kronecker's double series \cite{izm2002,Weil}, which are directly related to
elliptic $\theta$ functions.

Now we can easily write down all the terms of the exact asymptotic
expansion Eq. (\ref{Fnew}) of the free energy,  $F = - \ln Z$  for all models under consideration by using Eqs. (\ref{ExpansZ00}) and (\ref{ExpansZab}).

The bulk free energy $f_{\rm{bulk}}$ in Eq. (\ref{Fnew}) for the weighted spanning tree on finite $M \times N$ lattices for all boundary conditions is given by
\begin{eqnarray}
f_{\rm{bulk}}&=&-\frac{2}{\pi}\int_0^\pi\omega_z(x)dx =-\frac{1}{\pi}\sum_{n=0}^{\infty}(-1)^n(n+1/2)^{-2}z^{2n}
=-\frac{\Phi(-z^2,2,\frac{1}{2})}{\pi},
\label{fbulksp}
\end{eqnarray}
where $\Phi(-z^2,2,1/2)$ is the Lerch transcendent. In particular, for isotropic spanning tree ($z=1$), the Lerch transcendent is now $\Phi(-1,2,{1}/{2})=4 G$, where $G = 0.915 965 594 \dots$ is the Catalan constant. In what follow we can set $x_2=1$ without loss of generality.

\subsection{Spanning tree on the torus}
Using Eqs. (\ref{ZsptrTorFin}) and (\ref{ExpansZ00}), the exact asymptotic expansions of the free
energy for the spanning tree on torus, $F = - \ln Z_{M,N}^{Torus}$ can be written as
\begin{eqnarray}
F&=&-\ln Z_{M,N}^{Torus}=\ln S - 2\ln Z_{0,0}(z,M,N)\nonumber\\
&=& S f_{\rm{bulk}} -
\ln \xi- 4 \ln \eta(i z \xi)+4\pi\xi\sum_{p=1}^\infty \left(\frac{\pi^2
\xi}{S}\right)^p\frac{\Lambda_{2p}}{(2p)!} \frac{
K_{2p+2}^{0,0}(i z \xi)}{2p+2}, \label{ExpansZTorus}
\end{eqnarray}
where $f_{\rm{bulk}}$ is given by Eq. (\ref{fbulksp}). Thus the exact asymptotic expansion of the free energy for the periodic boundary conditions can be written in the form given by Eq. (\ref{Fnew}). The bulk free energy is given by Eq. (\ref{fbulksp}).
The surface free energy for the spanning tree $f_{1s}$ and $f_{2s}$ in Eq. (\ref{Fnew}) are equal to zero. For the leading correction terms $f_0(z\xi)$ we obtain
\begin{eqnarray}
f_0(z \xi)&=&-\ln{\xi}-4 \ln {\eta(i z\xi)}
\nonumber\\
&=&-2\ln {\eta(i z\xi)\eta(i/(z\xi))}+\ln{z},
\label{f0torsp}
\end{eqnarray}
in which  $\xi = \frac{N}{M}$.
Here we use the behavior of the Dedekind eta function $\eta(\tau')=\sqrt{-i \tau} \eta(\tau)$ under the Jacobi transformation $\tau \to \tau'=-1/\tau$,  for $\tau = i z \xi$.

For subleading correction terms $f_p(z \xi)$ for $p=1, 2, 3,\dots, $ we obtain
\begin{eqnarray}
f_p(z \xi)&=&4\pi^{2p+1} \xi^{p+1}\frac{\Lambda_{2p}}{(2p)!}
\frac{ K_{2p+2}^{0,0}(i z \xi)}{2p+2}. \nonumber
\end{eqnarray}
The coefficients $\Lambda_{2p}$ are listed in \cite{izm2014} and Kronecker's double series
$K_{2p+2}^{0,0}(i z \xi)$ in terms of the elliptic theta functions are given in \cite{izm2002,izm2003,izm2014} for
$p=1,2,3$ and $4$.

It is easy to see from Eq. (\ref{f0torsp}), that for the spanning tree on finite square lattices under periodic boundary conditions, $f_0(z\xi)$ does not contain the corner free energy $f_{\rm{corner}}$ given by Eq. (\ref{fcorner}), which confirm both conformal theory \cite{cardy1988} and finite-size scaling \cite{privman1988} predictions that logarithmic corner corrections to the free energy density should be absent for
periodic boundary conditions. However, such terms have been found by Duplantier
and David \cite{duplantier} in the two-dimensional spanning tree (ST) model under periodic boundary conditions
\begin{equation}
f_{\rm{corner}} = - \ln S \label{duplper}
\end{equation}
This discrepancy coming from the fact that Eq. (\ref{duplper}) has been obtained for the rooted spanning tree model.  The logarithmic correction to the free energy obtained by Duplantier and David (\ref{duplper}) is connected with the fact that number of rooted spanning trees is $S$ times larger than that of the un-rooted spanning trees (see Eq. (1.3) of \cite{duplantier}). It is not related to the contribution to free energy from the corner.
Taking into account that the result for the free energy for un-rooted spanning trees (considering in the present paper) differs from the rooted spanning trees by a factor $\ln S$, we can obtain the correct version for the corner free energy $f_{\rm{corner}}=0$ by adding to Eq. (\ref{duplper}) the term $\ln S$ .

\subsection{Spanning tree on the cylinder}

Using Eqs. (\ref{ZsptrCylFin}) and (\ref{ExpansZ00}), the exact asymptotic expansions of the free
energy for the spanning tree on cylinder, $F = - \ln Z_{M,N}^{Cyl}$ can be written as
\begin{eqnarray}
F&=&-\ln Z_{M,N}^{Cyl}=M{\rm arcsinh}\,1/z + \ln N+\ln z - \ln Z_{0,0}(z,M,2N)\nonumber\\
 &=&S f_{\rm{bulk}} - 2 \ln \eta(i z \xi)-\frac{1}{2}\ln 2
+\ln z-2\pi\xi\sum_{p=1}^\infty \left(\frac{\pi^2
\xi}{2S}\right)^p\frac{\Lambda_{2p}}{(2p)!} \frac{
K_{2p+2}^{0,0}(i z \xi)}{2p+2}. \label{ExpansZCyl}
\end{eqnarray}
Thus the exact asymptotic expansions of the free energy for the cylindrical boundary conditions can
be written in the form given by Eq. (\ref{Fnew}). The bulk free energy is given by Eq. (\ref{fbulksp}).
The surface free energy for the spanning tree $f_{1s}$ and $f_{2s}$ in Eq. (\ref{Fnew}) are
\begin{eqnarray}
f_{1s}&=&\frac{1}{2}\ln(z+\sqrt{1+z^2}),
\label{f1surfspc}\\
f_{2s}&=&0.
\label{f2surfspc}
\end{eqnarray}
For the leading correction terms $f_0(z\xi)$ we obtain
\begin{eqnarray}
f_0(z \xi)&=&-2\ln {\eta(i z\xi)}-\frac{1}{2}\ln{2}+\ln{z}
\nonumber\\
&=&-\ln {\eta(i z\xi)\eta(i/(z\xi))}-\frac{1}{2}\ln{2}+\frac{1}{2}\ln{z}-\frac{1}{2}\ln{\xi},
\label{f0freespc}
\end{eqnarray}
in which  $\xi = \frac{N}{M}$. For subleading correction
terms $f_p(z \xi)$ for $p=1, 2, 3, ..., $ we obtain
\begin{eqnarray}
f_p(z \xi)&=&\frac{\pi^{2p+1} \xi^{p+1}}{2^{p-1}}\frac{\Lambda_{2p}}{(2p)!}
\frac{ K_{2p+2}^{0,0}(i z \xi)}{2p+2}. \nonumber
\end{eqnarray}
The coefficients $\Lambda_{2p}$ are listed in \cite{izm2014} and Kronecker's double series
$K_{2p+2}^{0,0}(i z \xi)$ in terms of the elliptic theta functions are given in \cite{izm2002,izm2003,izm2014} for
$p=1,2,3$ and $4$.

\subsection{Spanning tree on the plane}

Using Eqs. (\ref{ZsptrFreeFin}) and (\ref{ExpansZ00}), the exact asymptotic expansions of the free
energy for the spanning tree on plane, $F = - \ln Z_{M,N}^{Free}$ can be written as
\begin{eqnarray}
F&=&-\ln Z_{M,N}^{Free}=\frac{1}{2}\ln S + N{\rm arcsinh}\,z + M{\rm arcsinh}\,1/z+\ln z
-\frac{1}{4}\ln(1+z^2)-\frac{1}{2}\ln 2  - \frac{1}{2}\ln Z_{0,0}(z,2M,2N)\nonumber\\
&=& S f_{\rm{bulk}} + N{\rm arcsinh}\,z + M{\rm arcsinh}\,1/z
+\frac{1}{4}\ln{S}-\frac{1}{4}\ln \xi- \ln \eta(i z \xi)+\ln z-\frac{1}{4}\ln(1+z^2)-\ln 2
\nonumber\\
&+&\pi\xi\sum_{p=1}^\infty \left(\frac{\pi^2
\xi}{4S}\right)^p\frac{\Lambda_{2p}}{(2p)!} \frac{
K_{2p+2}^{0,0}(i z \xi)}{2p+2}. \label{ExpansZFree}
\end{eqnarray}
Thus the exact asymptotic expansions of the free energy for the free boundary conditions can be
written in the form given by Eq. (\ref{Fnew}). The bulk free energy is given by Eq. (\ref{fbulksp}).
The surface free energy for the spanning tree $f_{1s}$ is given by Eq. (\ref{f1surfspc}) and $f_{2s}=-\frac{1}{2}\ln{z}+\frac{1}{2}\ln(1+\sqrt{1+z^2})$. For the leading correction terms $f_0(z\xi)$ we obtain
\begin{eqnarray}
f_0(z \xi)&=&\frac{1}{4}\ln{S}-\frac{1}{4}\ln{\xi}-\ln {\eta(i z\xi)}-\ln{2}-\frac{1}{4}\ln{(1+z^2)}
+\ln{z}
\nonumber\\
&=&\frac{1}{4}\ln{S}-\frac{1}{2}\ln {\eta(i z\xi)\eta(i/(z\xi))}-\ln{2}-\frac{1}{4}\ln{(1+z^2)}
+\frac{5}{4}\ln{z},
\label{f0freespf}
\end{eqnarray}
in which  $\xi = \frac{N}{M}$. For subleading correction terms $f_p(z \xi)$ for $p=1, 2, 3, ..., $ we obtain
\begin{eqnarray}
f_p(z \xi)&=&\frac{\pi^{2p+1} \xi^{p+1}}{2^{2p}}\frac{\Lambda_{2p}}{(2p)!}
\frac{ K_{2p+2}^{0,0}(i z \xi)}{2p+2}. \nonumber
\end{eqnarray}
The coefficients $\Lambda_{2p}$ are listed in \cite{izm2014} and Kronecker's double series
$K_{2p+2}^{0,0}(i z \xi)$ in terms of the elliptic theta functions are given in
\cite{izm2014} for $p=1,2,3$ and $4$.

It is easy to see from Eq. (\ref{f0freespf}) that for the spanning tree on finite square lattices under free boundary condition $f_0(z\xi)$ contains the universal part $f_{\rm{univ}}$ given by Eq. (\ref{funiv}). This confirms the conformal field theory prediction for the corner free energy in models for which the central charge is $c = - 2$. Moreover, $f_0(z\xi)$ contains the non-universal, geometry-independent constant $f_{\rm{nonuniv}}$ given by Eq. (\ref{fnu}).
Again, as in the case of periodic boundary conditions, there is discrepancy with the results of Duplantier and David \cite{duplantier} for the corner free energy in the spanning tree on finite square lattices under free boundary condition. They obtained for the corner free energy the expression
\begin{equation}
f_{\rm{corner}} =-\frac{3}{4} \ln S \label{duplper1}
\end{equation}
which is different from the conformal field theory prediction (\ref{fcorner}). Noting that the result for the un-rooted spanning tree differs from that of the rooted spanning tree by a factor $\ln S$, we obtain the correct version for the corner free energy given by Eq. (\ref{fcorner}) with $c = -2$ by adding to Eq. (\ref{duplper1}) the term $\ln S$.

\subsection{Spanning tree on the M\"{o}bius  strip}

Using Eqs. (\ref{ZsptrTorFin}) and (\ref{ExpansZ00}), the exact asymptotic expansions of the free
energy for the spanning tree on M\"{o}bius  strip, $F = - \ln Z_{M,N}^{Mob}$ can be written as
\begin{eqnarray}
F&=&-\ln Z_{M,N}^{Mob}=\ln M+N {\rm arcsinh}\,z - \ln Z_{0,0}(z,M,N)
- \ln Z_{\frac{1}{2},\frac{1}{2}}(z,M,N)\nonumber\\
&=& S f_{\rm{bulk}} + N{\rm
arcsinh}\,z-\ln \xi-\ln \theta_{\frac{1}{2},\frac{1}{2}}(i z
\xi)\eta(i
z\xi)\nonumber\\
&+&2\pi\xi\sum_{p=1}^\infty \left(\frac{\pi^2
\xi}{S}\right)^p\frac{\Lambda_{2p}}{(2p)!} \frac{ K_{2p+2}^{0,0}(i
z \xi)+K_{2p+2}^{\frac{1}{2},\frac{1}{2}}(i z \xi)}{2p+2}.
\label{ExpansZMob}
\end{eqnarray}
Thus the exact asymptotic expansions of the free energy for
M\"{o}bius  strip boundary conditions can be written in the form given by Eq. (\ref{Fnew}). The bulk free energy is given by Eq. (\ref{fbulksp}) and the surface free energy $f_{1s}$ is given by Eq. (\ref{f1surfspc}) and $f_{2s}$ in equal to zero.

For the leading correction terms $f_0(z\xi)$ we obtain
\begin{eqnarray}
f_0(z \xi)&=&-\ln{\xi}-\ln {\theta_{\frac{1}{2},\frac{1}{2}}(i z
\xi) \eta(i z\xi)}, \label{f0freespm}
\end{eqnarray}
in which  $\xi = \frac{N}{M}$. For subleading correction terms
$f_p(z \xi)$ for $p=1, 2, 3, ..., $ we obtain
\begin{eqnarray}
f_p(z \xi)&=&2\pi^{2p+1} \xi^{p+1}\frac{\Lambda_{2p}}{(2p)!}
\frac{ K_{2p+2}^{0,0}(i z
\xi)+K_{2p+2}^{\frac{1}{2},\frac{1}{2}}(i z \xi)}{2p+2}. \nonumber
\end{eqnarray}
The coefficients $\Lambda_{2p}$ are listed in \cite{izm2014}
and Kronecker's double series $K_{2p+2}^{0,0}(i z \xi)$ and
$K_{2p+2}^{\frac{1}{2},\frac{1}{2}}(i z \xi)$ in terms of the
elliptic theta functions are given in \cite{izm2002,izm2003,izm2014}.

\subsection{Spanning tree on the Klein bottle}

Using Eqs. (\ref{ZsptrTorFin}) and (\ref{ExpansZ00}), the exact asymptotic expansions of
the free energy for the spanning tree on the Klein bottle, $F = - \ln Z_{M,N}^{Klein}$ can be written as
\begin{eqnarray}
F&=&-\ln Z_{M,N}^{Klein}=\ln{2M} - \ln Z_{0,0}(z,M,2N)\nonumber\\
&=& S f_{\rm{bulk}} -\ln
2\xi-2\ln\eta(2i z\xi)+4\pi\xi\sum_{p=1}^\infty \left(\frac{\pi^2
\xi}{S}\right)^p\frac{\Lambda_{2p}}{(2p)!} \frac{ K_{2p+2}^{0,0}(2
i z \xi)}{2p+2}.
\label{ExpansZKlein}
\end{eqnarray}
Thus the exact asymptotic expansions of the free energy for the
Klein bottle boundary conditions can be written in the form given by
Eq. (\ref{Fnew}). The bulk free energy is given by Eq. (\ref{fbulksp}). The surface free energy $f_{1s}$ and $f_{2s}$ in Eq. (\ref{Fnew}) are equal to zero. For the leading correction terms $f_0(z\xi)$ we obtain
\begin{eqnarray}
f_0(z \xi)&=&-\ln{2\xi}-2 \ln {\eta(2i z\xi)}, \label{f0freespk}
\end{eqnarray}
in which  $\xi = \frac{N}{M}$.  For subleading correction terms
$f_p(z \xi)$ for $p=1, 2, 3, ..., $ we obtain
\begin{eqnarray}
f_p(z \xi)&=&4\pi^{2p+1} \xi^{p+1}\frac{\Lambda_{2p}}{(2p)!}
\frac{ K_{2p+2}^{0,0}(2i z \xi)}{2p+2}. \nonumber
\end{eqnarray}
The coefficients $\Lambda_{2p}$ are listed in \cite{izm2014}
and Kronecker's double series $K_{2p+2}^{0,0}(2 i z \xi)$ in terms
of the elliptic theta functions are given in \cite{izm2014} for $p=1,2,3$ and $4$.

\section{Conclusion}
\label{conclusion}

In this paper, we have used the method of \cite{izm2002} and \cite{izm2014} to derive exact finite-size corrections for the logarithm of the partition function  of the spanning-tree model on the $M \times N$ square lattice with five different sets of boundary conditions.
We have found that the exact asymptotic expansion of the free energy of the spanning-tree model can be written in the form given by Eq. (\ref{Fnew}). Except the bulk free energy $f_{bulk}$ all other coefficients in this expansion are sensitive to the boundary conditions. We have established several groups of new identities relating to the spanning-tree partition functions for different boundary conditions. We explain an apparent discrepancy between conformal field theory predictions and a two dimensional spanning tree model with periodic and free boundary conditions \cite{duplantier,brankov}. We have also obtained the corner free energy for  free boundary conditions. We proved the conformal field theory prediction about the corner free energy and have shown that the corner free energy, which is proportional to the central charge $c$, is indeed universal. We find the central charge in the framework of the conformal field theory to be $c=-2$.

\section{Acknowledgment}
\label{Acknowledgment}
This work were supported by a Marie Curie IIF (Project no. 300206-RAVEN)and IRSES
(Projects no. 295302-SPIDER and 612707-DIONICOS) within 7th European Community Framework Programme and
by the grant of the Science Committee of the Ministry of Science and Education of the Republic of
Armenia under contract 13-1C080.

\appendix

\section{Asymptotic expansion of $Z_{0,0}(z,M,N)$ and $Z_{\frac{1}{2}, \frac{1}{2}}(z,M,N)$}
\label{Expansion Z(0,0)(z,M,N)}
For the convenience of the reader,
in this appendix we  present the exact asymptotic expansions of
the logarithm of   $Z_{0, 0}(z,M,N)$ and $Z_{\frac{1}{2}, \frac{1}{2}}(z,M,N)$ given respectively
in Ref. \cite{izm2014} and Ref. \cite{izm2003}.

The exact asymptotic expansion of the logarithm of $Z_{0,0}(z,M,N)$ and $Z_{\frac{1}{2},\frac{1}{2}}(z,M,N)$ in terms of the Kronecker's double series \cite{izm2002,Weil} can be written as
\begin{eqnarray}
\ln Z_{0,0}(z,M,N)&=&\frac{S}{\pi}\int_0^{\pi} \omega_{z}(x)dx +
\ln\sqrt{S \xi}+ 2 \ln \eta(i z \xi)-2\pi\xi\sum_{p=1}^\infty \left(\frac{\pi^2
\xi}{S}\right)^p\frac{\Lambda_{2p}}{(2p)!} \frac{
K_{2p+2}^{0,0}(i z \xi)}{2p+2}, \label{ExpansZ00}\\
\ln Z_{\frac{1}{2},\frac{1}{2}}(z,M,N)&=&\frac{S}{\pi}\int_0^{\pi}
\omega_{z}(x)dx+
\ln \frac{\theta_{\frac{1}{2},\frac{1}{2}}(i z \xi)}
{\eta(i z \xi)}-2\pi\xi\sum_{p=1}^\infty
\left(\frac{\pi^2 \xi}{S}\right)^p\frac{\Lambda_{2p}}{(2p)!}
\frac{K_{2p+2}^{\frac{1}{2},\frac{1}{2}}(i z \xi)}{2p+2},
\label{ExpansZab}
\end{eqnarray}
where $S = M N$, $\xi = N/M$, $\eta(\tau)$ is the Dedekind - eta function, ${\rm K}_{2p}^{0,0}(\tau)$ and ${\rm
K}_{2p}^{\frac{1}{2},\frac{1}{2}}(\tau)$ is Kronecker's double series \cite{izm2002,Weil} and function $\theta_{\frac{1}{2},\frac{1}{2}}(i \tau)= \theta_{3}(\tau)$ is elliptic theta function.  $\Lambda_{2p}$ is the differential operators that have appeared here can be expressed via coefficients $z_{2p}$ of the Taylor expansion of the lattice dispersion relation $\omega_{z}(k)$ (see for example \cite{izm2014})
\begin{equation}
\omega_{z}(k)={\rm arcsinh}\left(z \sin{k}\right)=k\left(z +\sum _{p=1}^{\infty} \frac{z
_{2p}}{(2p)!}k^{2p}\right),
\label{SpectralFunction}
\end{equation}

The Kronecker's double series ${\rm K}_{2p}^{0,0}(\tau)$ and
${\rm K}_{2p}^{\frac{1}{2},\frac{1}{2}}(\tau)$ can all be expressed in terms of the elliptic
$\theta$-functions only \cite{izm2002,izm2003,izm2014}.

\end{document}